\documentclass[conference]{IEEEconf}
\input epsf
\usepackage{graphicx}

\usepackage{hyperref}

\usepackage{commands}
\usepackage{computed_values}

\begin{document}

\title{\textbf{\Large Follow Your Nose -- Which Code Smells are Worth Chasing?
\\}}

\author{Idan Amit$^{1}$, Nili Ben Ezra$^{1}$, and Dror G. Feitelson$^{1}$\\
	\normalsize $^{1}$The Hebrew University, Jerusalem, Israel\\
	\normalsize idan.amit@mail.huji.ac.il, feit@cs.huji.ac.il\\
}


\maketitle
\begin{abstract}
The common use case of static analysis alerts assumes causality:
Given an alert instance, fix it, and by doing so improve the code.
We empirically investigate  which alerts instances fixing improves quality and productivity.
We evaluated the alerts in \filesNum Java files from \reposNum GitHub repositories, all the Java repositories with 200+ commits in 2019.
We measured the influence of alerts on metrics for quality and productivity.
We present a list of properties that alerts should have if their removal is beneficial.
Out of 174 alerts computed by the CheckStyle static analyzer, less than 20\% were suitable, having these basic properties, and only a handful are rather robust.
The suitable alerts deal with simplicity, defensive programming, and abstraction.
Files without the suitable alerts are 50\% more likely to be of high quality.
\end{abstract}
\IEEEoverridecommandlockouts

\section{Introduction}
\label{sect:introduction}

High quality is desirable in software development.
A common method to improve quality is to identify low-quality code and improve it.
Alerts encapsulate software engineering knowledge and facilitate large-scale scanning of source code to identify potential problems \cite{arcellifontana13,7158503,1173068}.
Instance of an alert is a segment in the source that activated the logic.
\textbf{Alert instance fixing} is a refactoring of the code, keeping the same logic yet using an implementation in which the static analyzer no longer alerts.
Fixing the alert instance is believed to improve the quality of the code.
However, it seems that this approach suffers from many drawbacks.
Static analysis has been criticized for not representing problems of interest \cite{bavota15} and for having low predictive power \cite{10.1145/1368088.1368135, 5770619, kremenek2003z}.
Indeed, static analysis is not widely used by developers \cite{4602670, 6606613, do2020software, 10.1145/1368088.1368135, 8816730}.
Part of that is since alerts differ in their value.
Therefore, we would like to answer the following research questions.

\begin{quote}\emph{RQ: Which alert instance fixing improves the following software engineering goals: 1. fault proneness 2. productivity}\end{quote}

Note that we are not interested \textbf{only} in the predictive probability $P(C|A)$ (where $C$ is concept and $A$ is alert) but in the \emph{causal relation} \cite{pearl2009causality}
$P(Improve\:C|do(fix\: A))$, an alert whose instance fixing actually improves a goal.

The goals that we are interested in are quality and productivity, leading to a research question per goal.
Answering these research questions is important whether an alert fix is beneficial or not.
A small set of causal alerts, those whose instances fixing is beneficial,  can focus the refactor effort and efficiently improve the quality.
Non-causal alerts' instances can be untouched, saving wasted effort.

The common method to assess causality is controlled experiments.
Choose an alert, find instances where it occurs, fix them, and observe the results.
However, CheckStyle alone provides 174 alerts and assessing all of them will require tremendous effort.

We therefore use observations on the alerts'
effect `in nature' in order to identify the most promising alerts.
When looking for a causal relation from observations, one should identify a relation and see that it is not due to other variables.
We use properties that require predictive power of the alert with respect to the goal, and the existence of the relation also when \textbf{controlling the influential developer and length variables}.
We also use the change in variables over time and require co-change of the alert and goal.
Requiring all these attributes leaves us with a small set of alerts, benefiting programmers and allowing focused research.

We use corrective commit probability to measure fault proneness and commit duration to measure productivity.
The corrective commit probability (CCP) estimates the fraction of development effort that is invested in fixing bugs \cite{Amit2021CCP}.
Commit duration, the gross time to complete a commit, is an estimate of development effort \cite{Amit2021CCP}. It is computed as the average time from the previous commit of the same author on the same date, suitable to part time work.
Note that our target metrics are process metrics, unlike the alerts which are based on code, making dependency on a confounding variable less likely.

One could measure correlation of the alerts and the metrics, but correlation is not causation.
In a world of noisy observations, it is not easy to show causality.
On the other hand, it is possible to specify properties that are expected to hold if causality actually exists.
We suggest the following properties, required from a suitable alert:
\begin{itemize}
  \item \hyperref[sect:predictive-power]{Predictive power}: \cite{10.1145/1368088.1368135, 5770619, kremenek2003z} the ability to use alerts to detect low-quality files.
  \item \hyperref[sect:Monotonicity]{Monotonicity}: \cite{hill1965environment, swanson2015definition}  the alert should be more common in low-quality files than in high-quality files.
  \item \hyperref[sect:CoChange]{Co-change}: \cite{Amit2021CCP} an improvement in quality once an alert is fixed.
  \item \hyperref[sect:Twins]{Controlling \cite{pearl2009causality} the developer, influencing bug tendency}, comparing the quality of files of the same developer differing by an alert.
  \item \hyperref[sect:length-control]{Controlling \cite{pearl2009causality} length, influencing bug tendency}, to avoid alerts that are merely indications of large code volume.
\end{itemize}

The desired properties are derived from the alert fixing use case.
A causal alert should be predictive, more common in problematic files, beneficial when fixed, where the benefit is not due to the developer or length.

We used the `IllegalCatch' alert as a working example.
`IllegalCatch' identifies a too-wide exception class (e.g., Error, Exception) which therefore might catch and hide unintended \emph{other} exceptions.
It is a common best practice recommendation to avoid such catching \cite{lenarduzzi2020sonarqube,kery2016examining, bestPractices}.
We will show how our analysis supports this recommendation.

\subsection{Novelty and Contribution}

We suggest a novel method to evaluate if alert fixing is beneficial.
As far as we know, we are the first to investigate causality in alert fixing.
We applied the methodology for all CheckStyle's alerts, on a large dataset.
The results were that less than 20\% of the alerts were suitable, having all the desired properties with respect to a target metric.

We did this evaluation on metrics representing quality and productivity.
We also used a random metric as a negative control.
Our results agree with prior work on code review comments \cite{zampetti2022using}, influence on grades \cite{edwards2017investigating}, and on the importance of simplicity, abstraction, and defensive programming.
Our analysis also identifies infamous alerts like `IllegalToken' (go to) \cite{10.1145/362929.362947} and `InnerAssignment' (e.g., assignment in if statement) \cite{hovemeyer2004finding, edwards2017investigating, AssignmentVsEquality} as potentially causing bugs.
On the other hand, The predictive yet not causal `TodoComment' alert \cite{10.1145/3345629.3345631} is not identified as suitable.

We identify a small set of alerts that \textbf{might} be causal and therefore their fixing might improve the desired metric.
We show that files lacking them indeed have higher quality.
If these alerts are indeed causal, these are promising results for developers that will fix them.

\section{Related Work}
\label{sect:related-work}

Metrics are essential for quantitative research and therefore there is a long history of code-quality metrics research.
The need has motivated the development of several software quality models (e.g.\ \cite{boehm76,dromey95}).

\hide{For some, this term refers to the quality of the software product as perceived by its users \cite{schneidewind02}.
Capers Jones defined software quality as the combination of low defect rate and high user satisfaction \cite{Jones:1991:ASM:109758, CapersJonesQuality2012}.
Such a definition is not suitable for our needs since we do not have data about the users' satisfaction.
}
Code smells \cite{1999:RID:311424, 1173068} can be considered code quality metrics.
Yet, it is debatable whether they indeed reflect real quality problems \cite{alkilidar05,bavota15}.
One could treat each alert as a labeling function and evaluate them using their relations \cite{archimedes, alex2016data}.
Yet, this direction is not trivial and simpler alert-based methods might be circular.

Static analysis \cite{nagappan05, 4602670} is an analysis which is based on the source code, as opposed to dynamic analysis in which the code is being run.
We used the \href{https://checkstyle.sourceforge.io/}{CheckStyle} static analyser.
Alert instances are not necessarily problems but warnings.
There are no requirements from alerts and the developers implementing them choose their content.
While they capture knowledge, they are not always validated, a gap that we would like to fill.
Indeed, static analysis’s value was criticized \cite{ 10.1145/1368088.1368135, 5770619, kremenek2003z}.
It seems that sometimes developers do not: tend to use static analysis tools \cite{6606613}, choose refactoring targets based on static analysis, or consider alerts in pull request approval \cite{lenarduzzi2019does}.

Our work resembles that of Basili et al.\ \cite{544352} who validated the K\&C metrics \cite{Chidamber:1994:MSO:630808.631131} as predictors of fault-prone classes.
The difference is that we validate the alerts as potential causes and not as predictors.
Couto et al.\ \cite{couto2014predicting} used Granger causality tests, showing additional predictive power in time series for future events \cite{granger1969investigating}, in order to predict defects using code metrics.
Our work is also close in spirit to the work of Trautsch et al.\ that investigated the evolution of PMD static analyzer alerts and code quality, operationalized as defect density (defects per LOC) \cite{trautsch2020longitudinal}.
Another close research is of Ayewah et al.\, reporting on the experience of using the static analyzer FindBugs at Google \cite{ayewah2008using}.
Research on SonarCube's rules as fault proneness predictors showed that few rules are responsible for most benefit\cite{lomio2021fault, lenarduzzi2020sonarqube}.
Lenarduzzi et al. also found that the SonarCube rules similar to our suitable alerts `Hidden Field', `IllegalCatch' and `NestedIfDepth' are beneficial for bug identification \cite{lenarduzzi2020sonarqube}.

Edwards et al. investigated the relation between students' grades and alerts \textbf{categories} of CheckStyle and PMD \cite{edwards2017investigating}.
While there was no relation in general, there was a strong relation between coding flaw alerts (including our suitable alerts `Hidden Field', `FallThrough', and `InnerAssignment') and the grade.

Zampetti et al. evaluated CheckStyle alerts by their agreement with code review comments \cite{zampetti2022using}.
We report their results as an external validation of ours.
Note that in both code review and in grading a person recommends a change, fitting the causal use case.

\section{Methodology}

We discuss the \hyperref[sect:target-metrics]{metrics} that we are interested in improving.
We then define \hyperref[sect:Validation]{properties} that causal alerts should have.
Last, we construct a \hyperref[sect:dataset]{dataset} that enables us to evaluate the properties.

\subsection{Software Engineering Goals and Our Target Metrics}
\label{sect:target-metrics}

We aim to find the alerts whose fixing improves either quality (alerts' common use \cite{arcellifontana13,7158503,1173068}) or productivity.
Both quality and productivity are important software engineering goals.
They can be influenced by many aspects that are captured by alerts, like readability and complexity, hence there is a possible causal mechanism.

The target metrics that we use for the goals are corrective commit probability (CCP) for quality \cite{Amit2021CCP}.
We use commit duration for productivity.
Our main metric of interest is CCP since alerts' common use is to alert on possible bugs.
We use commit duration and the negative control in order to understand if alerts serve other goals too and as comparison.

A good metric should be reliable and valid.
A reliable metric should return similar results when measuring the same entity.
While in a physics lab one can repeat the same experiment multiple times, software projects evolve and we cannot measure the same entity repeatedly a few times.
Instead we measure the same metric on the same project in adjacent years and compare their values.
We then calculate the Pearson correlation, and the higher the value, the higher the stability of the project.
Quality and productivity have no agreed definition that we can validate with respect to it.
The metrics were validated in prior work \cite{Amit2021CCP} by agreement with developers referring to the goal concept (e.g. `low quality') and with other metrics of the same goal (e.g., Self Admitted Technical Debt for quality, commits per year for productivity).

It is also important to note that our metrics are process metrics, and not source code metrics.
If we used source code metrics, based on the same data source as alerts, that might lead to dependency between the alerts and the metrics which might bias the results.
Assuming conditional-independence given the concept (e.g., quality as concept) \cite{Blum:1998:CLU:279943.279962, 10.1007/BFb0026666} between the source code and process metrics, a decrease in quality is due to an increase of problems in the source code.
As far as we know, our metrics are the only validated \cite{Amit2021CCP} process metrics which are attributable to the file level for code quality and productivity.

\subsubsection{Corrective Commit Probability (CCP)}

Alerts are commonly used for defect prediction \cite{arcellifontana13,7158503,1173068}.
Defect fixes can be identified by linguistic analysis of the commit message.
The corrective commit probability (CCP) , which uses the intuition that bugs are bad, was defined and investigated in \cite{Amit2021CCP}.
CCP is the number of bug fix commits divided by the number of commits, a normalized commit bug fixed counting (maximum likelihood estimation is used on the project level in order to overcome linguistic analysis false negatives and positives ).

Projects with self-admitted technical-debt \cite{potdar2014exploratory}, developers referring to ``low quality'', and even swearing have a higher level of CCP than projects lacking them.
The Pearson correlation between the CCP values of the same project in adjacent years is 0.86, very stable \cite{Amit2021CCP}.

\subsubsection{Commit Duration}

It is common to measure productivity by output using (e.g., commits) per period.
In open source development such metrics are problematic since many  developers do not work full time hence the period is not estimated correctly.
We measure productivity using the average gross duration of a file's commits \cite{Amit2021CCP}, which is the inverse of output per period.
Therefore commit duration is correlated with output per period metrics yet more stable \cite{Amit2021CCP}.
Gross duration includes coffee breaks and lunch, which are part of the developer day of work.
We compute the gross duration as the time between a commit to the previous commit of the developer, \textbf{on the same day}.
The requirement for the same day overcomes large gaps of inactivity of open source developers that could be misinterpreted as long tasks.

The commit duration has 0.87 adjacent-years Pearson correlation, indicating that it is stable and reliable.
Commit duration co-changes with commits per developer and pull requests per developer, two common metrics for productivity.
In turn, commits-per-developer is correlated with self-rated productivity \cite{47853} and team lead perception of productivity \cite{Oliveira2020TLProd}.
Unlike commits per developer and pull requests per developer, commit duration can be associated with a file and therefore we choose it.
Smelly code is assumed to be harder to understand and modify, hurting productivity.

\subsubsection{Negative Control}

A negative control \cite{shi2020selective} is a metric that should be indifferent to the experiment.
Using this negative control, we can estimate the expected noise and evaluate the metrics of interest with respect to it.
As a negative control we used a random variable uniformly distributed between 1 and 100.
Table \ref{tab:smells-properties} shows that random, the negative control, has 3\% (4 alerts) suitable alerts, all with a hit rate 0.00 (rounded) and no robust alerts.

We chose to divide the metrics into groups of the 25\% `low-quality', 25\% `high-quality' and the `other' 50\% in the middle.
For all metrics, low values are `high-quality'.
The groups are large so we avoid problems due to imbalanced datasets \cite{van2007experimental, krawczyk2016learning, 10.1145/3338501.3357374}.
In the case of CCP, low-quality files are in the spirit of hot spots, files with many bugs, \cite{Tosun11Ai, Walkinshaw:2018:FRD:3239235.3239244, menzies2010defect, 5609530} and high-quality resemble files of reduced-risk \cite{Niedermayr18Trivial}.

The target metrics implementation has many small details, listed in the supplementary materials \cite{replication}.

\subsection{Desired Properties of Causal Alerts}
\label{sect:Validation}

We list here properties that causal alerts \textbf{should have}.
We call alerts that have these properties \textbf{suitable}.
We \textbf{do not claim that suitable alerts are causal}.
Non causal alerts might have the discussed properties due to many reasons, further discussed.

We claim that a causal relation should have the properties that we used.
Predictive power, monotonicity, and co-change are common causality properties, regardless of the domain \cite{10.1145/1368088.1368135, 5770619, kremenek2003z, imtiaz2019developers, hill1965environment, swanson2015definition, Amit2021CCP}.
Controlling variables are also very common \cite{pearl2009causality}.
We choose the developer and file length due to their large influence (See Figures \ref{fig:ccp_by_line_count_deciles}, \ref{fig:developer_deciles_by_ccp}), which should be controlled.
Hence a causal relation should indicate low quality, lower quality when it is more common and when it increases, and predictive ability, regardless of the developer or the file length.

Though alerts might be valuable due to their predictive power alone, their use implicitly assumes causality.
That since alerts' common use is ``once you find an alert instance, fix it, in order to avoid the problems it causes''.

A common approach in causality studies is to build models of possible relations among all variables \cite{pearl2009causality}.
However, alerts’ usage assumes that an alert instance is important on its own, regardless of other instances and context.
We do the same and treat each alert separately, aiming for finding an alert-concept relation and not relations between alerts.
This allows us to use properties that examine each alert on its own and show the expected temporal behavior.
Note that the separate analysis poses a threat of identifying a non causal alert as causal .
A `TodoComment' added and removed with `IllegalCatch' will seem causal though it is merely predictive.

The direction of causality, if it exists, is known.
Given concept $C$ and alert $A$, one could build models for $A$ causes $ C$ and for $C$ causes $A$, and see which is better supported.
However, we know that a alert in the source code might lead to a bug,
but the bugs will not modify the source code, invalidating $C$ causes $A$.
Due to these assumptions we are left with a simple model for causality of  alerts and concepts.

\begin{figure}[!ht]
\centering
\includegraphics[width=0.48\textwidth,trim={0mm 0mm 0 0mm},scale=0.3]{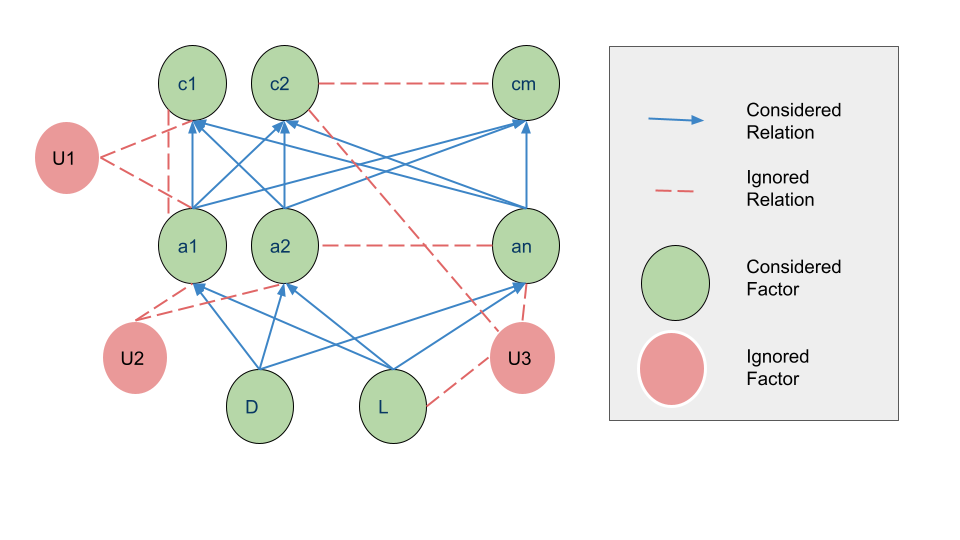}
\caption{\label{fig:Causalty_Invetigated_Model}
The full arrows indicate the considered relations. The dashed lines indicate examples of \emph{unconsidered relations}. $a_{i}$ - alert, $c_{j}$ - concept, \textbf{D}eveloper,  \textbf{L}ength, and \textbf{U}nknown.
}
\end{figure}

Figure \ref{fig:Causalty_Invetigated_Model} summarizes the graphical model that we use.
We consider relations between an alert $a_{i}$ and a concept $c_{j}$.
We take into account the \textbf{D}eveloper and \textbf{L}ength.
We do not consider the influence of a concept on an alert, influence of other \textbf{U}nknown factors, relations between two alerts, relations between two concepts, and relations between multiple such entities.

Note that extending the model will make it more complex and require more data to evaluate.
Adding more requirements will reduce the number of suitable alerts and the current model already shows that the fixing of the vast majority of the alerts is not likely to be beneficial.

While we do not explicitly address some factors, like programming language and problem domain, they are automatically controlled in comparison of the same file on different dates.

We suggest a \emph{basic} set of properties that should hold given causality.
The properties are \hyperref[sect:predictive-power]{predictive power}, \hyperref[sect:Monotonicity]{monotonicity}, \hyperref[sect:CoChange]{co-change of alert and target metric},  \hyperref[sect:Twins]{influence when the developer is controlled}, and \hyperref[sect:length-control]{influence when the length is controlled}.

By removing alerts that do not have these properties, we obtain a group of less than 20\% of the 151 original alerts that can be further investigated for causality.
The rest of the alerts should be treated with more suspicion, with respect to causality.
However, these alerts might be useful for other uses.
For example, features with predictive power can guide testing.

\subsubsection{Predictive Power}
\label{sect:predictive-power}

We start with requiring predictive power.
Precision, defined as $P(positive|hit)$, measures a classifier's tendency to avoid false positives.
In a software development context, this measures the probability of actually finding a low-quality file when checking an alert instance.

Precision might be high simply since the positive rate is high.
Precision lift, defined as
\begin{equation}
Precision/P(positive) - 1 = \frac{P(positive|hit) - P(positive)}{P(positive)}
\end{equation}
, copes with this difficulty and measures the additional probability of having a true positive relative to the base positive rate.
For example, a useless random classifier will have precision equal to the positive rate, but a precision lift of 0.

There are additional supervised learning metrics \hyperref[sect:robust-smells-influence]{that we discuss below}.
However, the precision lift best captures the usage of alerts and therefore it is a metric of high interest.
When a developer looks for low-quality code to improve, the precision lift measures the value of using the alert as a guide rather than going over the source code randomly.
High precision and precision lift make the use of alerts effective.

We define the predictive power property to be the conjunction of three terms:
\begin{itemize}
\item
We require a positive precision lift in order to claim that observing the hits of an alert is better than observing random files. This is deliberately the minimal threshold of improvement.
\item
We also require that the target metric average given an alert will be higher than the metric average.
Note that this measures \emph{how bad} the code is, and not only whether it is bad.
\item
We required that the alert appears in
200+ files since
a small number of cases might make the evaluation inaccurate.
Other than that, a rare alert provides less opportunities for quality improvement.
\end{itemize}

Our working example, `IllegalCatch' has precision of 0.32 with respect to CCP, meaning that a third of the files with this alert are of low quality.
The mean CCP of files with `IllegalCatch' is 0.25, meaning that a quarter of the commits modifying these files are bug fixes.

\subsubsection{Monotonicity}
\label{sect:Monotonicity}

In case that an alert causes more problems, we expect monotonicity, more alerts in problematic files \cite{hill1965environment, swanson2015definition}.
Hence the alert hit rate in `high-quality' files will be lower than in `other' files, which in turn will be lower than in `low-quality' files.

Here is the justification.
Let $L$ be `low-quality' and let $A$ be alert.
If $A$ causes $L$, then
$P(L|A) > P(L) > P(L|\neg A)$.
Using Bayes rule
\begin{equation}\label{eq:bayes-rule}
P(L|A) = \frac{P(A|L) \cdot P(L)}{P(A)}
\end{equation}
the inequality $P(L|A) > P(L)$ can be rewritten
\begin{equation}
\frac{P(A|L) \cdot P(L)}{P(A)} > P(L)
\end{equation}
, which implies $P(A|L) > P(A)$ as desired.
The same derivation for no alert leads to $P(\neg A|L) < P(\neg A)$.

Extending to three groups, if we compare the alert appearance probability in `high-quality', `other', and `low-quality' files we expect it to be strictly monotonically increasing.
`IllegalCatch',our example, is indeed monotone with respect to quality.

\subsubsection{Co-Change with Target Metric}
\label{sect:CoChange}

In causality we would like to investigate the influence of the ``do'' action \cite{pearl2009causality}.
Since we have observations, we use co-change analysis \cite{Amit2021CCP}, analysing observations of change.
We compare the change of two metrics on the same entity in two consecutive periods.
As with correlation, co-change of two metrics does not necessarily imply causality.
However, causal relations do imply co-change, since a change in the causing metric should lead to a change in the affected metric in some contexts.
Co-change analysis was used in software engineering to show that coupling improvement co-changes with CCP improvement \cite{Amit2021CCP}.

\newcommand{\imp}[1]{\mbox{$m_{#1}\!\!\uparrow$}}
In short, denote the event that metric $i$ improved from one year to the next by \imp{i}.
The probability $P(\imp{j} \: | \: \imp{i})$ (the equivalent to precision in supervised learning) indicates how likely we are to observe an improvement in metric $j$ knowing of an improvement in metric $i$.
Alert improvement is defined as a reduction in the alert occurrences relative to the same file the year before.
Hence, we actually investigate observations of $P(Improve\:C|do(fix\: A))$.
Target metric improvement is a reduction in its value between the years.
It might be that we will observe high precision but it will be simply since $P(\imp{j})$ is high.
In order to exclude this possibility, we also observe the precision lift,
$\frac{P(\imp{j} \: | \: \imp{i})}{P(\imp{j})} -1$.
Note that these are the same precision and precision lift used in the predictive power, this time applied to temporal data.
Note that though we investigate $P(Improve\:C|do(fix\: A))$, there might be more variables $X$ influencing $P(Improve\:C|do(fix\: A), X)$ the behavior and therefore co-change alone is not enough to demonstrate causality.

`IllegalCatch' has precision lift of 15\% in co-change, and improves with CCP more than expected had they been unrelated.

\subsubsection{Influence when Controlling the Developer}
\label{sect:Twins}

\begin{figure}[!ht]
\centering
\includegraphics[width=0.48\textwidth,trim={0mm 0mm 0 0mm},clip]{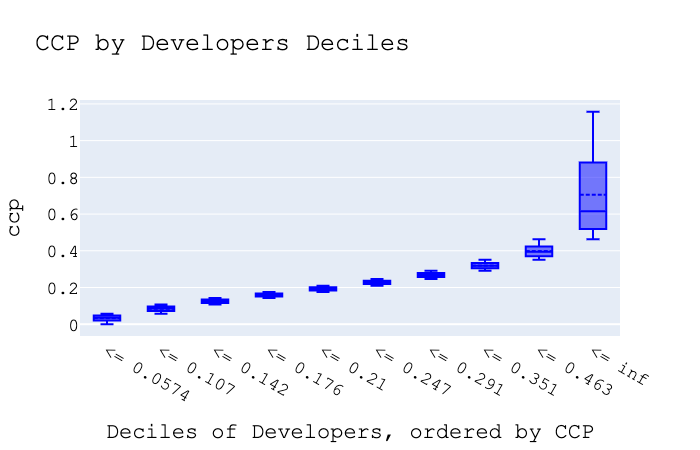}
\caption{\label{fig:developer_deciles_by_ccp}
The average CCP of median developers is very far from those in the extremes.
}
\end{figure}

Figure \ref{fig:developer_deciles_by_ccp} shows that the average CCP of developers is very diverse, with 6\% for top 10\% developers, compared to a median of 21\%.
We use variable control \cite{pearl2009causality} in order to avoid a benefit that is not due to the alert fixing but due to the developer behavior.
The developer writing the code influences it in many ways.
Consider an alert that has no influence from a software engineering point of view, yet due to sociological reasons it divides the developer community into two groups.
The reader might think of the spaces vs. tabs wars as a possible example.
If one of the groups tends to produce code of higher quality, it will look as if this alert influences the quality.

The statistical method to avoid indirect influence is to control the relevant variables:
Investigate $P(Q|A,C)$ instead of $P(Q|A)$, where $Q$ is quality, $A$ is alert, and $C$ is community.
Note that in case that we condition on a community that has no influence then $P(Q|A,C)=P(Q|A)$, and this property will not be influential.

However, we do not have information about the community or any other confounding variables.
A popular solution in psychology is ``twin experiments'' \cite{vandenberg1966contributions}.
Identical twins have the same biological background, so a difference in their behavior is attributed to another difference (e.g., being raised differently).

Twin experiments thus help to factor out personal style, biases, and skill, and can neutralize differences in the developer community and other influencing factors.

In this paper we consider files written by the same developer in the same repository as twins.
That is that the same developer is being factored out, leaving differences not related to the developer, like the alerts in the file.

`IllegalCatch' has 37\% precision lift in twin experiments, hence files with this alert tend significantly to have more bugs than other files of the same developer.

\subsubsection{Influence when Controlling File Length}
\label{sect:length-control}

\begin{figure}[!ht]
\centering
\includegraphics[width=0.48\textwidth,trim={0mm 0mm 0 0mm},clip]{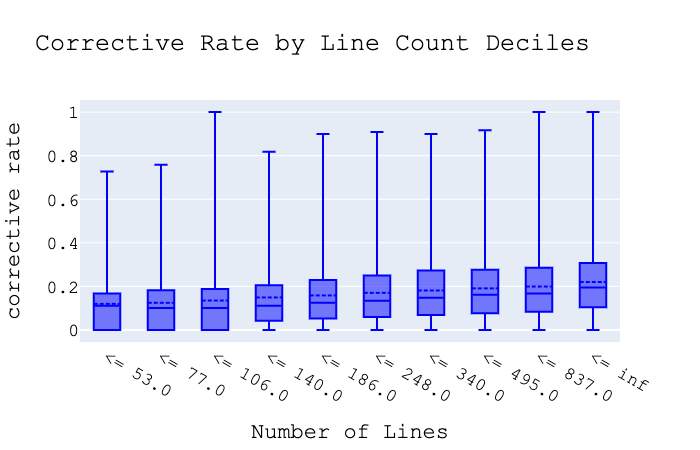}
\caption{\label{fig:ccp_by_line_count_deciles}
Median in solid, mean in dashed. Corrective commit ratio without MLE estimation of CCP since number of file commits is small.
}
\end{figure}

The influence of length on error-proneness is one of the most established results of software engineering \cite{lipow1982number, gil17}.
As Figure \ref{fig:ccp_by_line_count_deciles} shows, file of at most 104 lines, the first 3 deciles, have average CCP less than 0.1, which is enough to enter the top 20\% of projects \cite{Amit2021CCP}.

Doing a co-change analysis of line count and
CCP, an improvement in line count has a 5\% precision lift
predicting CCP improvement, and 6\% precision lift predicting
a significant CCP improvement of 10 percentage points.

Hence, the appearance of an alert instance might seem indicative of low quality simply since the instance is indicative of length, which in turn is indicative of lower quality \cite{gil17}.
In order to avoid that, we divide files into length groups, and require the property of predictive power when the length group is controlled \cite{pearl2009causality}.
Specifically, we require precision lift in the short (first 25\%, up to \shortLengthThreshold lines), medium (25\% to 75\%), and long (highest 25\%, at least \longLengthThreshold lines) length groups.

Also, we filter out alerts whose Pearson correlation with line count is higher than 0.5, which is \href{https://www.statisticssolutions.com/pearsons-correlation-coefficient/}{considered to indicate high correlation}.
This should have been a relatively easy requirement since the \href{https://checkstyle.sourceforge.io/config_metrics.html#JavaNCSS}{`JavaNCSS'} alert (number of Java statements above a threshold) has Pearson correlation of 0.45 and the \href{https://checkstyle.sourceforge.io/config_sizes.html#FileLength}{`FileLength'} alert (length higher than a threshold) has a Pearson correlation of 0.44 with the line count.
These alerts are basically step functions on length and have lower than 0.5 correlation.
Hence, 0.5 correlation is a strong correlation also in our specific case.
Indeed, 127 alerts have lower than 0.5 Pearson correlation with length.

`IllegalCatch' has precision lift of 16\% for short, 15\% for medium, and 12\% for long files.
Hence, relation to length is small and the higher lift for shorter files might be to their higher average quality (See same behavior in Section \ref{seq:groups}).

\subsection{Dataset Creation}
\label{sect:dataset}

Our dataset starts with all \href{https://console.cloud.google.com/bigquery?d=github\_repos\&p=bigquery-public-data&page=dataset}{BigQuery GitHub schema} non-redundant%
\footnote{Excluding forks and other related repositories.}
Java repositories with 200+ commits during 2019, taken from Amit and Feitelson \cite{Amit2021CCP}.
There are 909 such repositories.
From these we extract all non-test Java files with at least 10 commits during 2019 that we could analyze with \href{https://checkstyle.sourceforge.io/}{CheckStyle}.
This led to \filesNum files from \reposNum repositories.

From each commit we extracted the files involved in it, the commit message and author.
Process metrics were computed on all file commits in a year.
Metrics computation is explained in Section \ref{sect:target-metrics} and related work.
Basically, CCP uses a linguistic model to identify fixes and compute their probability.
Commit duration is the average time between the commit and the previous commit of the same author, computed only on two commits in the same day to avoid weekends and time off.

For the predictive power analysis, we used commits of 2019 and the code version of January 1st 2019.
For temporal analysis (stability, co-change, alert fixing), we used the commits of 2017, 2018, and 2019 the code versions of January 1st of the relevant year.

The computation of static analysis based dataset is long, so research on it is usually based on a single tool \cite{trautsch2020longitudinal, ayewah2008using, lomio2021fault, lenarduzzi2020sonarqube, zampetti2022using}.
Source code tends to deviate from language specifications, and static analyzers might fail due to such deviations \cite{10.1145/1646353.1646374}.
Therefore running CheckStyle requires a lot of tuning and monitoring.
This technical constraint also limited extending the dataset further into the past.

We used CheckStyle to analyze the code and identify alerts.
\textbf{The name CheckStyle is misleading}; it is a general purpose static analyzer, having stylistic alerts but also alerts on coding, design, complexity, etc.
We chose CheckStyle since it is popular \cite{7476667, zampetti2017open, vassallo2020developers}, open-source, with many diverse alerts, and of precision 86\%, way above others \cite{lenarduzzi2021critical}.
In addition, its alerts tend to be general and not specific low-level erroneous patterns.
Due to the dataset creation effort we could not use more than one static analyzer.
This tool currently recognizes 174 alerts, 151 appeared 200+ times and are relevant to us.

We excluded test files since their behavior is different from that of functional code.
We identified test files by looking for the string `test' in the file path.
Berger at el.\ reported 100\% precision of this classifier, based on a sample of 100 files \cite{DBLP:journals/corr/abs-1901-10220}.

\section{Results}

We present statistics on alerts that appeared in 200+ non-test files.
Our research questions are being answered by presenting a table of alerts for each metric.
The suitable alerts with respect to CCP are presented in Table \ref{tab:code_smell_influence_CCP}, and commit duration in Table \ref{tab:code_smell_influence_Duration}.
We analyze \filesNum files that had at least 10 commits during 2019, from \reposNum repositories.

Table \ref{tab:smells-properties} summarizes how many alerts have each property with respect to each target metric.
We perform sensitivity analysis, by increasing and reducing the required threshold and observing the change in the number of resulting alerts.
In the 'Robust' setting we require a precision lift of 10\% in predictive power, twins, length and co-change.
In the 'Almost' setting we allow a small negative -10\% precision lift.

\begin {table*}[h!]\centering
\caption{ \label{tab:smells-properties} Percentage of Alerts with Each of the Properties }
\begin{tabular}{ | l| l| l| l| l| l| l| l| l| l  | }
\hline
\textbf{Concept} & \textbf{Description} & \textbf{Suitable} & \textbf{Robust} & \textbf{Almost} & \textbf{Predictive} & \textbf{Cochange} & \textbf{Developer} & \textbf{Monotonicity} & \textbf{Length} \\
\hline
CCP & Fix commit probability & 0.18 & 0.03 & 0.41 & 0.70 & 0.74 & 0.75 & 0.59 & 0.35\\ \hline
Duration & Avg. commit duration  & 0.13 & 0.01 & 0.41 & 0.79 & 0.64 & 0.75 & 0.57 & 0.34\\ \hline
Random & Negative control & 0.03 & 0.00 & 0.27 & 0.32 & 0.41 & 0.48 & 0.20 & 0.21\\ \hline
\end{tabular}
\end{table*}

Most alerts are designed to warn on potential bugs, hence most are expected to be aligned with CCP.
We observed good alignment with productivity too, as many alerts warn on aspects like readability (e.g., `BooleanExpressionComplexity', `OneStatementPerLine'), which influence productivity.
The negative control has few alerts, presenting the level of noise.

Our work is novel so there is no similar benchmark that we can evaluate our work directly against.
We evaluated the results based on these criteria.
First, the alert content and its relation to the metric.
Prior work on the suitable alerts is additional support.
We observe groups of alerts, since over or under representation of a group is less likely to be accidental.
In the same manner, we compare the amount of result to the negative control, representing the level of noise.
An alert being suitable for more than a single metric is another indication of its value.
The `Robust' group, presenting higher than 10\% lift indicates a stronger influence.
Last, we compare our results with the agreement with code review comments as another external validation.

\subsection{Evaluating Alerts Influence}
\label{sect:robust-smells-influence}

Supervised learning metrics shed light on the common behavior of a classifier and a concept.
However, a metric should be used with respect to a need.
We explain to which software engineering needs each metric that we use here corresponds.

The cases in which the concept is true are called `positives' and the positive rate is denoted $P(positive)$.
Cases in which the classifier is true are called `hits' and the hit rate is $P(hit)$.
A high hit rate of an alert means that the developer will have to examine many files.

Ideally, hits correspond to positives but usually they differ.
Precision, discussed above, is one example of a metric for such differences.

Recall, defined as $P(hit|positive)$, measures how many of the positives are also hits; in our case, this is how many of the low quality files an alert identifies.
Recall is of high importance to developers who aim to fix all low quality.

Accuracy, $P(positive=hit)$ is probably the most common supervised learning metric.
However, it is misleading when there is a large group of null records, samples that are neither hits nor positives.
For example, the vast majority of programs are benign, so always predicting benign will have high accuracy but detect no malware \cite{10.1145/3338501.3357374}.
True Positives (TP, correct hits), False Positives (FP, wrong hits), and False Negatives (FN, unidentified positives) are the cases in which either the concept or the classifier are true---without the null records.
The Jaccard index \cite{real1996probabilistic} copes with the problem of null records by omitting them, computing $\frac{TP}{TP+FP+FN}$.

Tables \ref{tab:code_smell_influence_CCP} and
\ref{tab:code_smell_influence_Duration} present the above statistics for suitable alerts, ordered by precision with the robust alerts in bold.
The concept is low-quality, and the precision, recall, and Jaccard are computed with respect to this concept.
The mean (of the target metric) is not a supervised learning metric.
It does not measure if the quality is low but how low it is (the higher the value the worse, in our target metrics).
Co-change and Twins present the precision lift in these properties.
Removal probability is the probability of an alert existing in a file in one year to be removed in the year after it.
The removal probability is not part of the causality properties but a measure of the tendency of developers to act upon alerts.

The strongest result observed is the poor performance of even the suitable alerts.
The highest precision is less than 50\%, implying at least 1 mistake per successful identification, and the common results are around 30\%, not much more than the 25\% expected from a random guess.
Such performance quickly leads to mistrust.

The hit rate of most alerts is low, possibly since developers already consider them to be a bad practice and avoid them in the first place.

The recall and the Jaccard index, aggregating the precision and recall, is also very low for all alerts.
Thus, these alerts are of limited value either for developers looking for easy wins (precision oriented) or those aiming to win the war (recall oriented).

\begin {table*}[h!]\centering
\caption{ \label{tab:code_smell_influence_CCP} Suitable Alerts for CCP (ordered by precision, robust alerts in bold)}
\begin{tabular}{ | l| l| l| l| l| l| l| l| l| l| }
\hline
  &   &   &   & \textbf{Hit} &   &   & \textbf{Co-} &  & \textbf{Removal}\\
\textbf{Alert} & \textbf{CheckStyle Group} & \textbf{Precis.} & \textbf{Mean} & \textbf{Rate} & \textbf{Recall} & \textbf{Jaccard} & \textbf{change} & \textbf{Twins} & \textbf{Prob.}\\
\hline
NestedTryDepth & Coding & 0.39 & 0.32 & 0.01 & 0.02 & 0.02 & 0.01 & 0.24 & 0.07\\ \hline
FallThrough & Coding & 0.39 & 0.29 & 0.01 & 0.01 & 0.01 & 0.16 & 0.54 & 0.11\\ \hline
EmptyForIteratorPad & Whitespace & 0.39 & 0.31 & 0.01 & 0.02 & 0.01 & 0.05 & 0.43 & 0.09\\ \hline
InnerAssignment & Coding & 0.35 & 0.28 & 0.03 & 0.04 & 0.04 & 0.01 & 0.34 & 0.05\\ \hline
IllegalThrows & Coding & 0.34 & 0.24 & 0.00 & 0.00 & 0.00 & 0.16 & 0.93 & 0.09\\ \hline
\textbf{ParameterAssignment} & Coding & 0.34 & 0.28 & 0.11 & 0.14 & 0.11 & 0.13 & 0.32 & 0.05\\ \hline
\textbf{NPathComplexity} & Metrics & 0.34 & 0.27 & 0.17 & 0.22 & 0.16 & 0.18 & 0.27 & 0.05\\ \hline
MethodParamPad & Whitespace & 0.34 & 0.26 & 0.02 & 0.03 & 0.03 & 0.30 & 0.28 & 0.14\\ \hline
AnonInnerLength & Size Violation & 0.33 & 0.27 & 0.07 & 0.09 & 0.07 & 0.31 & 0.48 & 0.09\\ \hline
\textbf{UnnecessaryParentheses} & Coding & 0.33 & 0.26 & 0.13 & 0.16 & 0.12 & 0.21 & 0.34 & 0.08\\ \hline
NestedIfDepth & Coding & 0.33 & 0.27 & 0.20 & 0.25 & 0.17 & 0.24 & 0.31 & 0.04\\ \hline
\textbf{IllegalCatch} & Coding & 0.32 & 0.25 & 0.20 & 0.25 & 0.16 & 0.15 & 0.37 & 0.04\\ \hline
JavaNCSS & Metrics & 0.32 & 0.27 & 0.15 & 0.19 & 0.13 & 0.23 & 0.26 & 0.05\\ \hline
\textbf{AvoidStaticImport} & Import & 0.32 & 0.26 & 0.21 & 0.26 & 0.17 & 0.16 & 0.31 & 0.04\\ \hline
RequireThis & Coding & 0.32 & 0.32 & 0.00 & 0.00 & 0.00 & 0.45 & 0.28 & 0.13\\ \hline
ClassDataAbstractionCoupling & Metrics & 0.32 & 0.26 & 0.19 & 0.23 & 0.16 & 0.12 & 0.36 & 0.04\\ \hline
ExecutableStatementCount & Size Violation & 0.32 & 0.26 & 0.20 & 0.25 & 0.16 & 0.20 & 0.25 & 0.04\\ \hline
JavadocParagraph & JavaDoc & 0.32 & 0.25 & 0.19 & 0.23 & 0.16 & 0.09 & 0.37 & 0.10\\ \hline
TrailingComment & Misc & 0.32 & 0.26 & 0.22 & 0.26 & 0.17 & 0.30 & 0.33 & 0.05\\ \hline
VisibilityModifier & Class Design & 0.31 & 0.25 & 0.23 & 0.27 & 0.17 & 0.14 & 0.12 & 0.05\\ \hline
VariableDeclarationUsageDistance & Coding & 0.31 & 0.26 & 0.10 & 0.12 & 0.09 & 0.21 & 0.19 & 0.09\\ \hline
IllegalToken & Coding & 0.31 & 0.27 & 0.01 & 0.01 & 0.01 & 0.98 & 0.43 & 0.03\\ \hline
ExplicitInitialization & Coding & 0.30 & 0.24 & 0.13 & 0.15 & 0.11 & 0.18 & 0.21 & 0.05\\ \hline
HiddenField & Coding & 0.30 & 0.23 & 0.46 & 0.52 & 0.23 & 0.18 & 0.32 & 0.03\\ \hline
EqualsHashCode & Coding & 0.29 & 0.22 & 0.00 & 0.00 & 0.00 & 1.02 & 0.33 & 0.37\\ \hline
AvoidStarImport & Import & 0.29 & 0.24 & 0.15 & 0.17 & 0.12 & 0.06 & 0.27 & 0.08\\ \hline
WhitespaceAround & Whitespace & 0.28 & 0.24 & 0.24 & 0.26 & 0.16 & 0.09 & 0.20 & 0.13\\ \hline
\end{tabular}
\end{table*}

\begin {table*}[h!]\centering
\caption{ \label{tab:code_smell_influence_Duration} Suitable Alerts for Duration }
\begin{tabular}{ | l| l| l| l| l| l| l| l| l| l| }
\hline
  &   &   & \textbf{Mean}  & \textbf{Hit} &   &   & \textbf{Co-} &   & \textbf{Removal}\\
\textbf{Alert} & \textbf{CheckStyle Group} & \textbf{Precis.} & \textbf{(minutes)} & \textbf{Rate} & \textbf{Recall} & \textbf{Jaccard} & \textbf{change} & \textbf{Twins} & \textbf{Prob.}\\
\hline
InterfaceTypeParameterName & Naming & 0.48 & 146.09 & 0.00 & 0.00 & 0.00 & 1.00 & 0.60 & 0.04\\ \hline
\textbf{AvoidEscapedUnicodeCharacters} & Misc & 0.46 & 137.42 & 0.01 & 0.01 & 0.01 & 0.24 & 0.26 & 0.05\\ \hline
ArrayTypeStyle & Misc & 0.40 & 132.44 & 0.02 & 0.03 & 0.03 & 0.08 & 0.28 & 0.10\\ \hline
FileLength & Size Violation & 0.40 & 134.45 & 0.02 & 0.04 & 0.04 & 0.11 & 0.14 & 0.02\\ \hline
MethodTypeParameterName & Naming & 0.37 & 152.63 & 0.00 & 0.00 & 0.00 & 0.11 & 0.13 & 0.04\\ \hline
BooleanExpressionComplexity & Metrics & 0.36 & 126.56 & 0.05 & 0.07 & 0.06 & 0.12 & 0.12 & 0.06\\ \hline
EmptyForInitializerPad & Whitespace & 0.36 & 136.80 & 0.00 & 0.00 & 0.00 & 0.33 & 0.10 & 0.43\\ \hline
EmptyCatchBlock & Block & 0.35 & 118.11 & 0.03 & 0.05 & 0.04 & 0.19 & 0.21 & 0.11\\ \hline
FallThrough & Coding & 0.35 & 120.35 & 0.01 & 0.01 & 0.01 & 0.09 & 0.26 & 0.11\\ \hline
UnnecessarySemicolonInEnumeration & Coding & 0.33 & 133.25 & 0.01 & 0.01 & 0.01 & 0.27 & 0.09 & 0.08\\ \hline
IllegalImport & Import & 0.33 & 121.88 & 0.00 & 0.00 & 0.00 & 0.72 & 0.25 & 0.33\\ \hline
NestedForDepth & Coding & 0.33 & 119.86 & 0.02 & 0.03 & 0.03 & 0.12 & 0.13 & 0.08\\ \hline
NPathComplexity & Metrics & 0.32 & 118.79 & 0.17 & 0.23 & 0.15 & 0.01 & 0.10 & 0.05\\ \hline
OneStatementPerLine & Coding & 0.32 & 122.85 & 0.01 & 0.01 & 0.01 & 0.16 & 0.06 & 0.11\\ \hline
ParameterNumber & Size Violation & 0.32 & 116.64 & 0.08 & 0.10 & 0.08 & 0.11 & 0.22 & 0.05\\ \hline
JavadocParagraph & JavaDoc & 0.31 & 113.84 & 0.19 & 0.24 & 0.16 & 0.05 & 0.15 & 0.08\\ \hline
TrailingComment & Misc & 0.31 & 117.03 & 0.22 & 0.27 & 0.17 & 0.06 & 0.06 & 0.05\\ \hline
ClassTypeParameterName & Naming & 0.31 & 120.42 & 0.00 & 0.00 & 0.00 & 0.50 & 0.64 & 0.02\\ \hline
NoFinalizer & Coding & 0.29 & 109.10 & 0.00 & 0.00 & 0.00 & 0.43 & 0.49 & 0.13\\ \hline
WriteTag & JavaDoc & 0.27 & 104.99 & 0.55 & 0.60 & 0.23 & 0.01 & 0.18 & 0.02\\ \hline
\end{tabular}
\end{table*}

\subsection{Suitable Alerts Investigation}

In this section we investigate the resulting suitable alerts, potentially causal, and some alerts surprisingly not considered as such.
The descriptions are based on the \href{https://checkstyle.sourceforge.io/checks.html}{CheckStyle tutorial}.

\subsubsection{Suitable Alerts for CCP}

CheckStyle groups the alerts by type, which allows us to analyze the findings with respect to these groups.
For CCP, Table \ref{tab:code_smell_influence_CCP}
shows that many of the alerts are `coding' checks, typically alerting on a local specific case.
This is aligned with student's lower grades in tasks with coding alerts but not with alerts in general \cite{edwards2017investigating}.
The `metrics' alerts are more general, representing simplicity and abstraction.
The `size' group also represents simplicity while its member `AnonInnerLength' also checks abstraction.
The alerts `class-design' and `VisibilityModifier' check abstraction too.
The `import' and the `illegal' groups (subsets of the coding group) check that a too wide action is not used.
When the alerts group were compared to code review comments `coding' had precision of 60.8\%, `size' 56\%, `Whitespace' 55.6\%, `class-design' of 39.4\%, `imports' 20.8\%
, while `metrics', `misc' and `illegal' were not reported \cite{zampetti2022using}.
Methods are too far for quantitative comparison yet the groups represented in our analysis have high precision.

While they do not change the code behavior, human programmers might change their behavior due to alerts like `whitespace', `JavaDoc comments', and `misc'.
Refactoring of the same spirit was shown to be useful \cite{10.1145/3345629.3345631}.
It might be due to their influence on the developer, like in improving the readability.
Note that since we use twin experiments, it is possible that part of their predictive power is due to indication about the developer's behavior, yet more influence exists.

It is interesting that entire groups are not represented at all, questioning their utility.
Most of them also have low precision with code review comments \cite{zampetti2022using}: `header' (87.5\%), `block' (47.2\%), `naming conventions’ (23.7\%), `annotations' (0\%), `unnecessary',  and `regexp' groups.

\subsubsection{Suitable Alerts for Commit Duration}

Duration suitable alerts
(Table \ref{tab:code_smell_influence_Duration}) come mainly from `coding', `metrics', `naming conventions' and `size violation'.
Yet it is more diverse than the CCP groups and the indirect groups are more represented.
There are 20 suitable alerts, much more than those related to the negative control.

\subsubsection{Stronger Suitable Alerts}
Few alerts are suitable for both target metrics, indicating extra validation of their usefulness.

\href{https://checkstyle.sourceforge.io/config\_metrics.html#NPathComplexity}{`NPathComplexity'} \cite{10.1145/42372.42379} was designed based on cyclomatic complexity \cite{mccabe76}, checking the number of execution paths, is potential for CCP and duration.
So is \href{https://checkstyle.sourceforge.io/config_coding.html#FallThrough}{`FallThrough'}, dropping in switch statement options without a break.

\href{https://checkstyle.sourceforge.io/config_javadoc.html#JavadocParagraph}{`JavadocParagraph'}, a JavaDoc formatting alert,  and \href{https://checkstyle.sourceforge.io/config_misc.html#TrailingComment}{`TrailingComment'}, a comment in a code line,  are also suitable for CCP and commit duration.
They seem not harmful on their own but indicative of deviation from a standard.

\href{https://checkstyle.sourceforge.io/config_javadoc.html#JavadocParagraph}{'JavadocParagraph'}, a JavaDoc formatting alert,  and \href{https://checkstyle.sourceforge.io/config_misc.html#TrailingComment}{'TrailingComment'}, a comment in a code line,  are also suitable for CCP and commit duration.
They seem not harmful on their own but indicative of deviation from a standard.

More alerts of special interest are the `Robust' ones, which have lift higher than 10\%.
The robust alerts for CCP are `IllegalCatch' (our working example) \cite{kery2016examining, bestPractices, lenarduzzi2020sonarqube, lenarduzzi2020sonarqube}, `AvoidStaticImport',   `ParameterAssignment', `UnnecessaryParentheses', and the already discussed `NPathComplexity' \cite{10.1145/42372.42379}.
Compared to code review comments \cite{zampetti2022using} `IllegalCatch' has precision of 76.9\%, `UnnecessaryParentheses' of 71.4\%,  `AvoidStaticImport' of only 12.5\% and the rest were not reported.

Defensive programming \cite{10.5555/948835.948838} is programming in a way that is robust to \emph{future programming mistakes}, and not avoiding only incorrectness in the current implementation.
`AvoidStaticImport' allows hiding the relation between a static method and its source (e.g., `import static java.lang.Math.pow;' and calling pow instead of Math.pow), which is a source of confusion.
\href{https://checkstyle.sourceforge.io/config_coding.html#ParameterAssignment}{`ParameterAssignment'} disallows assignment of parameters.
\href{https://checkstyle.sourceforge.io/config_coding.html#IllegalCatch}{`IllegalCatch'} alerts on the use of a too general class in catch (e.g., Error, Exception).
Besides the robust defensive alerts, pay attention to
\href{https://checkstyle.sourceforge.io/config_coding.html#IllegalToken}{`IllegalToken'}, a structured variant of `go to', the most notorious construct in software engineering \cite{dijkstra1968letters}.
Also, \href{https://checkstyle.sourceforge.io/config_coding.html#InnerAssignment}{`InnerAssignment'}  \cite{hovemeyer2004finding} ``Checks for assignments in subexpressions, such as in String s = Integer.toString(i = 2);'', is causing bugs since the early days of C.
`UnnecessaryParentheses' is harmless on its own but might indicate misunderstanding of the developer.

The only robust alert with respect to duration is `AvoidEscapedUnicodeCharacters', which advocates a readable representation of Unicode.
That might hurt understanding the text.


Another interesting aspect is that some of the alerts consider abstraction: `ClassDataAbstractionCoupling'
, `VisibilityModifier', and `AnonInnerLength'.

\subsubsection{Non Suitable Alerts and Sensitivity Analysis}
\label{sect:surprising}

There were alerts that we were surprised by their failure to be suitable for CCP.
In the easier to understand cases, some alerts did not have a positive precision lift when conditioned on a certain length group value.
These are the classic `CyclomaticComplexity' \cite{mccabe76} (ancestor of NPathComplexity)  and `MagicNumber'.
This might be accidental since we conducted many experiments and some alerts have a low hit rate and therefore are sensitive to noise.
It is also possible that an alert is indeed beneficial but only on a certain length group.

Harder cases are alerts that do not have the predictive or co-change property.
This is more disturbing since these properties are the core assumptions in alerts usage.
We were surprised that the alerts `BooleanExpressionComplexity' and `ConstantName' did not have a positive co-change precision lift, and `SimplifyBooleanExpression' had a negative predictive precision lift.

In order to evaluate the sensitivity of the properties, we present in Table \ref{tab:smells-properties} alerts that are `Robust' and `Almost' alerts.
Even with the generous `Almost' setting, about two thirds of the alerts are not included.
Requiring robust properties leads to only a handful of alerts.

Another way to perform sensitivity analysis is to count the number of properties alerts have.
This way we know how close the alert is to have all properties.
The CCP suitable alerts, 27 of them, have all 5 properties.
49 alerts have 4 properties, 31 alerts have 3 properties, 31 alerts have 2 properties, 13 alerts have 1 property and 2 have none.

Out of the alerts having 4 properties, 73\% lack the length property, 12\% lack the co-change property, 10\% lack the twins, 4\% lack the monotonicity, and none lack predictive power.

\subsection{Predictive Power of Alerts Groups}
\label{seq:groups}

We checked if the combined power of all the suitable alerts for a metric is higher than each of them alone.
For each metric, we checked the probability of being either `high-quality' or `low-quality', given that none of the suitable alerts appear in the file (see Table \ref{tab:group_smell_influence}).
Note that in this section we no longer investigate causality but modeling, as we \emph{observe} the alerts’ existence.
If the alerts are causal, their removal is expected to lead to what we observe.

For CCP the gap is very high, and the probability of being high-quality is 44\%, with a lift of 50\%.
Note that the high-quality CCP means no bugs (at least in 10 commits, real probability might be somewhat higher), making the result outstanding.
When we control for length the precision lift is only 1\% for the already good short files, 32\% for medium, and 59\% for long.

The gap for duration is quite small.

\begin {table}[h!]\centering
\caption{ \label{tab:group_smell_influence} Predictive Power of Alerts Groups}
\begin{tabular}{ | l| l| l| l  | }
\hline
\textbf{Metric} & \textbf{Hit Rate} & \textbf{High Quality} & \textbf{Low Quality} \\
\hline
CCP & 0.15 & 0.44 & 0.24\\ \hline
Duration & 0.27 & 0.25 & 0.22\\ \hline
\end{tabular}
\end{table}

For the predictive analysis of the alerts we used data from 2019.
In order to avoid evaluation on the same data we used 2018 in this section.
Yet, we warn of data leakage.
Some of it is due to co-change that uses prior years data.
More leakage comes from the stability of the alerts and the metrics.
The file CCP has an adjacent-year Pearson correlation of 0.45, and most alerts are even more stable, with an average Pearson correlation of 0.93.
This stability means that the train and test samples are similar.


\section{Threats to Validity}

We conducted a focused research, raising threats of external validity.
Our dataset contains only Java projects, simplifying its construction and taking language differences out of the equation.
But this raises a question whether the results hold for other languages too.

We used a single static analyser, \href{https://checkstyle.sourceforge.io/}{CheckStyle}.
Other tools (e.g., FindBugs \cite{10.1145/1646353.1646374}, SonarCube \cite{campbell2013sonarqube}, Coverity \cite{10.1145/1646353.1646374, imtiaz2019developers}, PMD \cite{trautsch2020longitudinal}, \href{https://www.perforce.com/products/klocwork}{Klocwork}, and DECOR \cite{10.1109/TSE.2009.50}) offer different alerts, some of which might be causal.
Note that not using more tools does not invalidate the results obtained on CheckStyle, they mean that CheckStyle alerts are only part of the available ones.

The dataset contains only files on which we could run CheckStyle.
Using uncommon syntax can make static analyzers fail \cite{10.1145/1646353.1646374}, so one should not assume that the omitted files behave like those in our dataset.
We also restricted our files to be files with at least 10 commits.
We did so since estimating the quality based on less commits will be very noisy.
While 812 projects had such files, these files are only 26\% of the files.
It is possible that in files in which the number of commits is small, the behavior of alerts is different.

One could use other target metrics or even aim for different software engineering goals, obtaining different results, and identifying different alerts of interest.
We used three software engineering metrics in order not to be governed by a single one.

Comparison to similar research increases the validity of our results.
Yet, even the close in spirit work of Lomio et al. \cite{lomio2021fault} that analyzes the predictive power of SonarCube rules for fault proneness is not comparable simply since the underlying rules are different.
We showed that our results are aligned with code review remarks \cite{zampetti2022using}, student grades  \cite{edwards2017investigating}, and prior work on specific alerts \cite{zampetti2022using, dijkstra1968letters, hovemeyer2004finding,edwards2017investigating, 10.1145/42372.42379, kery2016examining, lenarduzzi2020sonarqube},  all providing external validation.


Alert instances fixing are not always small atomic actions.
Commits might be tangled \cite{6624018, herbold2020largescale} and serve more than one goal.
Even when serving one goal, an alert might be removed as part of a large clean up, along with more influential alerts.
It is also possible that while one alert was removed, a different alert was introduced.
In principle, this can be taken care of by controlling all other alerts.
However, controlling all other alerts will be too restrictive in a small dataset, leading to basing the statistics on very few cases.
In the same spirit, while we controlled length and the developer, there might be other confounding variables influencing the results.

We analyzed the source code on a specific date and used the commits done in an entire year in order to measure quality.
It is possible that we identified an alert in January, and a developer fixed it in February, making it irrelevant for the rest of the year.
But alerts removal probability is low (e.g., \cite{HAMDI2021106699}, Table \ref{tab:code_smell_influence_CCP}), indicating that this scenario is rare.

Our classification of files into `low-quality', `high-quality' and `other’ are necessarily noisy.
As an illustrative example of the implication of a small number of commits, consider a file with CCP of 0.1, a probability of 10\% that a commit will be a bug fix.
The file has a probability of 35\% to have no hit in 10 commits and it will be mis-classified as `high-quality'.

On top of that, the VC dimension theory \cite{doi:10.1137/1116025, vapnik2013nature} warns us that given a fixed size dataset, the more statistics we compute, the more likely that some of them will deviate from the true underlying probabilities.
We started with 151 alerts and for each one of them we computed five properties for three target metrics, relying on even more statistics.
This might lead to missing causal alerts or misidentifying an alert as causal.
Note that an alert has a low probability to achieve all properties with respect to the negative control.
On the other hand, doing a lot of statistical tests, some will fail.
Some causal alerts that have the desired properties might accidentally fail on our data.
Examining these properties on more files will help to further validate the results.

Another threat comes from the use of the linguistic model in order to identify corrective actions.
The linguistic model accuracy is 93\%, close to that of human annotators \cite{Amit2021CCP}.
Yet, a misclassification of a commit as a bug fix might change a file quality group.

Some properties that are not related to quality, like bug detection efficiency and programming language, influence CCP.
For example, as predicted by Linus's law \cite{LinusRule}, popular projects with more eyeballs detect bugs more efficiently and have higher CCP \cite{Amit2021CCP}.
However, the tests' presence has an adjacent year Pearson of 0.83 and time from previous touch, a lower bound on time to detect a bug has an adjunct year Pearson of 0.51.
Also, the detection efficiency influence is small relative to the difference in CCP between projects, which can reach a factor of 6 or more.
Besides, we have relative properties like co-change and twins where the project properties are fixed.
Therefore these properties are robust to this influence.

The suitability of the properties to our needs pose a threat on the suitability of the results.
Aiming for causality existence, we required positive influence and ignored the effect size.
One can argue that an alert with high effect size is interesting even if it lacks other properties.
We also examined overall precision, neglecting the possibility of conditional benefit.
For example, defensive programming alerts might be valuable if a mistake is made in the future.
However, if this mistake would not be done, the value of the defensive act will not be captured in our properties.

\section{Future Work}

There are endless process metrics that removal of alerts might improve like security and run time performance.
We provide our dataset and the code that can enable such research.


Low-quality files lacking important alerts are hints for developing new alerts, capturing hitherto unrecognized problems.
Defect prediction is usually based on alerts, and alerts are usually hand-crafted by experts.
Given the dataset of high and low quality files, one can try an ``end-to-end'' approach \cite{amit2021end, choetkiertikul2016deep} and build models whose input is the source code itself.

We took a large step from mere correlation, yet we did not prove a causal relation between alerts and the target metrics.
Our dataset provides intervention opportunities.
One can choose an alert of interest and retrieve its instances.
Then deliberately modify the code to fix the instance, either by a suggestion to the author or directly.
Such experiments are the classical causality testbed and given our data they can be focused and easier to conduct.
In such experiments we will know that the removal targets were not chosen due to a confounding variable that also influences the quality.
Avoiding complex modifications will remove the threat that the alert removal was a by-product of a complex change, changing the quality due to different reasons.

\section{Conclusions}

Identification of causes for low quality is an important step in quality improvement.
We showed that many alerts are due to code that does not seem to cause low quality.
We did identify alerts that have properties expected from low quality causes.

For researchers, we presented a methodology for identifying possibly causal variables.
Out of 151 candidate alerts, we found that less than 20\% have our properties and may have a causal relationship with quality.

For developers, we first justify their mistrust of most alerts.
We also suggest a small set of interesting alerts and basic software engineering guidelines: simplicity, defensive programming, and abstraction.
We suggest using CheckStyle to identify the suitable alerts and actively remove them, starting in the longer files since short ones are usually of better quality.
That will save the vast majority of effort invested in other alerts (27 suitable, or even 5 robust, instead of 174 alerts), make the problem detection effort more efficient, and is likely to improve the code quality.

Files that have no suitable CCP alerts have 44\% probability to have no bugs and show improvement even when the length is controlled.
While we still did not prove causality, the results indicate a stronger relation between these alerts and improved quality, making the decision to act upon them more justifiable.

\label{sec:Supplementary}
All supplementary materials can be found at  \cite{replication}.
\anon{
Analysis used the infrastructure for GitHub \cite{Amit2021General}, natural language processing for software engineering  \cite{Amit2021CommitClassification}, and analysis utilities \cite{Amit2021Analysis}.
All other supplementary materials can be found at \url{https://github.com/evidencebp/follow-your-nose}.}{}

\anon{
\section*{Acknowledgements}

This research was supported by the ISRAEL SCIENCE FOUNDATION (grant No.\ 832/18).
We thank Udi Lavi, Daniel Shir, Yinnon Meshi, Tamar Litvak, and Millo Avissar for their help, discussions and insights.
}{}

%
\IEEEpeerreviewmaketitle

\bibliographystyle{abbrv}
\bibliography{abbrv.bib,bibtex.bib}

\begin{thebibliography}{10}

\bibitem{alkilidar05}
H.~Al-Kilidar, K.~Cox, and B.~Kitchenham.
\newblock The use and usefulness of the {ISO/IEC} 9126 quality standard.
\newblock In {\em Intl.\ Synp.\ Empirical Softw.\ Eng.}, pages 126--132, Nov
  2005.

\bibitem{Amit2021Analysis}
I.~Amit.
\newblock Analysis utilities.
\newblock Aug 2021.

\bibitem{amit2021end}
I.~Amit.
\newblock End to end software engineering research.
\newblock {\em arXiv preprint arXiv:2112.11858}, 2021.

\bibitem{Amit2021General}
I.~Amit.
\newblock General infrastructure for github data analysis over bigquery.
\newblock Aug 2021.

\bibitem{Amit2021CommitClassification}
I.~Amit.
\newblock natural language processing for software engineering.
\newblock Aug 2021.

\bibitem{10.1145/3345629.3345631}
I.~Amit and D.~G. Feitelson.
\newblock Which refactoring reduces bug rate?
\newblock In {\em Proceedings of the Fifteenth International Conference on
  Predictive Models and Data Analytics in Software Engineering}, PROMISE’19,
  page 12–15, New York, NY, USA, 2019. Association for Computing Machinery.

\bibitem{Amit2021CCP}
I.~Amit and D.~G. Feitelson.
\newblock Corrective commit probability: a measure of the effort invested in
  bug fixing.
\newblock {\em Software Quality Journal}, pages 1--45, Aug 2021.

\bibitem{archimedes}
I.~Amit, E.~Firstenberg, and Y.~Meshi.
\newblock Framework for semi-supervised learning when no labeled data is given.
\newblock U.S. patent application \#US20190164086A1, 2017.

\bibitem{arcellifontana13}
F.~Arcelli~Fontana, V.~Ferme, A.~Marino, B.~Walter, and P.~Martenka.
\newblock Investigating the impact of code smells on system's quality: An
  empirical study on systems of different application domains.
\newblock pages 260--269, 09 2013.

\bibitem{4602670}
N.~{Ayewah}, W.~{Pugh}, D.~{Hovemeyer}, J.~D. {Morgenthaler}, and J.~{Penix}.
\newblock Using static analysis to find bugs.
\newblock {\em IEEE Software}, 25(5):22--29, 2008.

\bibitem{ayewah2008using}
N.~Ayewah, W.~Pugh, D.~Hovemeyer, J.~D. Morgenthaler, and J.~Penix.
\newblock Using static analysis to find bugs.
\newblock {\em IEEE software}, 25(5):22--29, 2008.

\bibitem{544352}
V.~R. {Basili}, L.~C. {Briand}, and W.~L. {Melo}.
\newblock A validation of object-oriented design metrics as quality indicators.
\newblock {\em IEEE Transactions on Software Engineering}, 22(10):751--761,
  1996.

\bibitem{bavota15}
G.~Bavota, A.~De~Lucia, M.~Di~Penta, R.~Oliveto, and F.~Palomba.
\newblock An experimental investigation on the innate relationship between
  quality and refactoring.
\newblock {\em J.\ Syst.\ \& Softw.}, 107:1--14, Sep 2015.

\bibitem{7476667}
M.~{Beller}, R.~{Bholanath}, S.~{McIntosh}, and A.~{Zaidman}.
\newblock Analyzing the state of static analysis: A large-scale evaluation in
  open source software.
\newblock In {\em 2016 IEEE 23rd International Conference on Software Analysis,
  Evolution, and Reengineering (SANER)}, volume~1, pages 470--481, 2016.

\bibitem{DBLP:journals/corr/abs-1901-10220}
E.~D. Berger, C.~Hollenbeck, P.~Maj, O.~Vitek, and J.~Vitek.
\newblock On the impact of programming languages on code quality.
\newblock {\em CoRR}, abs/1901.10220, 2019.

\bibitem{10.1145/1646353.1646374}
A.~Bessey, K.~Block, B.~Chelf, A.~Chou, B.~Fulton, S.~Hallem, C.~Henri-Gros,
  A.~Kamsky, S.~McPeak, and D.~Engler.
\newblock A few billion lines of code later: Using static analysis to find bugs
  in the real world.
\newblock {\em Commun. ACM}, 53(2):66–75, Feb 2010.

\bibitem{Blum:1998:CLU:279943.279962}
A.~Blum and T.~Mitchell.
\newblock Combining labeled and unlabeled data with co-training.
\newblock In {\em Proceedings of the Eleventh Annual Conference on
  Computational Learning Theory}, COLT' 98, pages 92--100, New York, NY, USA,
  1998. ACM.

\bibitem{boehm76}
B.~W. Boehm, J.~R. Brown, and M.~Lipow.
\newblock Quantitative evaluation of software quality.
\newblock In {\em Intl.\ Conf.\ Softw.\ Eng.}, number~2, pages 592--605, Oct
  1976.

\bibitem{AssignmentVsEquality}
K.~L. Busbee.
\newblock Assignment vs equality.

\bibitem{campbell2013sonarqube}
G.~Campbell and P.~P. Papapetrou.
\newblock {\em SonarQube in action}.
\newblock Manning Publications Co., 2013.

\bibitem{Chidamber:1994:MSO:630808.631131}
S.~R. Chidamber and C.~F. Kemerer.
\newblock A metrics suite for object oriented design.
\newblock {\em IEEE Trans. Softw. Eng.}, 20(6):476--493, Jun 1994.

\bibitem{choetkiertikul2016deep}
M.~Choetkiertikul, H.~K. Dam, T.~Tran, T.~Pham, A.~Ghose, and T.~Menzies.
\newblock A deep learning model for estimating story points, 2016.

\bibitem{couto2014predicting}
C.~Couto, P.~Pires, M.~T. Valente, R.~S. Bigonha, and N.~Anquetil.
\newblock Predicting software defects with causality tests.
\newblock {\em Journal of Systems and Software}, 93:24--41, 2014.

\bibitem{10.1145/362929.362947}
E.~W. Dijkstra.
\newblock Letters to the editor: Go to statement considered harmful.
\newblock {\em Commun. ACM}, 11(3):147–148, Mar 1968.

\bibitem{dijkstra1968letters}
E.~W. Dijkstra.
\newblock Letters to the editor: go to statement considered harmful.
\newblock {\em Communications of the ACM}, 11(3):147--148, 1968.

\bibitem{do2020software}
L.~N.~Q. Do, J.~Wright, and K.~Ali.
\newblock Why do software developers use static analysis tools? a user-centered
  study of developer needs and motivations.
\newblock {\em IEEE Trans.\ Softw.\ Eng.}, 2020.

\bibitem{dromey95}
G.~Dromey.
\newblock A model for software product quality.
\newblock {\em IEEE Trans.\ Softw.\ Eng.}, 21(2):146--162, Feb 1995.

\bibitem{edwards2017investigating}
S.~H. Edwards, N.~Kandru, and M.~B. Rajagopal.
\newblock Investigating static analysis errors in student java programs.
\newblock In {\em Proceedings of the 2017 ACM Conference on International
  Computing Education Research}, pages 65--73, 2017.

\bibitem{1999:RID:311424}
M.~Fowler.
\newblock {\em Refactoring: Improving the Design of Existing Code}.
\newblock Addison-Wesley Longman Publishing Co., Inc., Boston, MA, USA, 1999.

\bibitem{gil17}
Y.~Gil and G.~Lalouche.
\newblock On the correlation between size and metric validity.
\newblock {\em Empirical Softw.\ Eng.}, 22(5):2585--2611, Oct 2017.

\bibitem{granger1969investigating}
C.~W. Granger.
\newblock Investigating causal relations by econometric models and
  cross-spectral methods.
\newblock {\em Econometrica: journal of the Econometric Society}, pages
  424--438, 1969.

\bibitem{HAMDI2021106699}
O.~Hamdi, A.~Ouni, M.~O. Cinneide, and M.~W. Mkaouer.
\newblock A longitudinal study of the impact of refactoring in android
  applications.
\newblock {\em Information and Software Technology}, 140:106699, 2021.

\bibitem{herbold2020largescale}
S.~Herbold, A.~Trautsch, B.~Ledel, A.~Aghamohammadi, T.~A. Ghaleb, K.~K.
  Chahal, T.~Bossenmaier, B.~Nagaria, P.~Makedonski, M.~N. Ahmadabadi,
  K.~Szabados, H.~Spieker, M.~Madeja, N.~Hoy, V.~Lenarduzzi, S.~Wang,
  G.~Rodríguez-Pérez, R.~Colomo-Palacios, R.~Verdecchia, P.~Singh, Y.~Qin,
  D.~Chakroborti, W.~Davis, V.~Walunj, H.~Wu, D.~Marcilio, O.~Alam, A.~Aldaeej,
  I.~Amit, B.~Turhan, S.~Eismann, A.-K. Wickert, I.~Malavolta, M.~Sulir,
  F.~Fard, A.~Z. Henley, S.~Kourtzanidis, E.~Tuzun, C.~Treude, S.~M. Shamasbi,
  I.~Pashchenko, M.~Wyrich, J.~Davis, A.~Serebrenik, E.~Albrecht, E.~U. Aktas,
  D.~Strüber, and J.~Erbel.
\newblock Large-scale manual validation of bug fixing commits: A fine-grained
  analysis of tangling.
\newblock {\em arXiv preprint arXiv:2011.06244}, 2020.

\bibitem{6624018}
K.~{Herzig} and A.~{Zeller}.
\newblock The impact of tangled code changes.
\newblock In {\em 2013 10th Working Conference on Mining Software Repositories
  (MSR)}, pages 121--130, 2013.

\bibitem{hill1965environment}
A.~B. Hill.
\newblock The environment and disease: association or causation?, 1965.

\bibitem{hovemeyer2004finding}
D.~Hovemeyer and W.~Pugh.
\newblock Finding bugs is easy.
\newblock {\em Acm sigplan notices}, 39(12):92--106, 2004.

\bibitem{imtiaz2019developers}
N.~Imtiaz, B.~Murphy, and L.~Williams.
\newblock How do developers act on static analysis alerts? an empirical study
  of coverity usage.
\newblock In {\em 2019 IEEE 30th International Symposium on Software
  Reliability Engineering (ISSRE)}, pages 323--333. IEEE, 2019.

\bibitem{8816730}
N.~{Imtiaz}, A.~{Rahman}, E.~{Farhana}, and L.~{Williams}.
\newblock Challenges with responding to static analysis tool alerts.
\newblock In {\em 2019 IEEE/ACM 16th International Conference on Mining
  Software Repositories (MSR)}, pages 245--249, 2019.

\bibitem{bestPractices}
T.~Janssen.
\newblock 9 best practices to handle exceptions in java, July 2017.

\bibitem{6606613}
B.~{Johnson}, Y.~{Song}, E.~{Murphy-Hill}, and R.~{Bowdidge}.
\newblock Why don't software developers use static analysis tools to find bugs?
\newblock In {\em 2013 35th International Conference on Software Engineering
  (ICSE)}, pages 672--681, 2013.

\bibitem{5609530}
Y.~Kamei, S.~Matsumoto, A.~Monden, K.~i.~Matsumoto, B.~Adams, and A.~E. Hassan.
\newblock Revisiting common bug prediction findings using effort-aware models.
\newblock In {\em 2010 IEEE International Conference on Software Maintenance},
  pages 1--10, Sept 2010.

\bibitem{kery2016examining}
M.~B. Kery, C.~Le~Goues, and B.~A. Myers.
\newblock Examining programmer practices for locally handling exceptions.
\newblock In {\em 2016 IEEE/ACM 13th Working Conference on Mining Software
  Repositories (MSR)}, pages 484--487. IEEE, 2016.

\bibitem{krawczyk2016learning}
B.~Krawczyk.
\newblock Learning from imbalanced data: open challenges and future directions.
\newblock {\em Progress in Artificial Intelligence}, 5(4):221--232, 2016.

\bibitem{kremenek2003z}
T.~Kremenek and D.~Engler.
\newblock Z-ranking: Using statistical analysis to counter the impact of static
  analysis approximations.
\newblock In {\em International Static Analysis Symposium}, pages 295--315.
  Springer, 2003.

\bibitem{lenarduzzi2020sonarqube}
V.~Lenarduzzi, F.~Lomio, H.~Huttunen, and D.~Taibi.
\newblock Are sonarqube rules inducing bugs?
\newblock In {\em 2020 IEEE 27th International Conference on Software Analysis,
  Evolution and Reengineering (SANER)}, pages 501--511. IEEE, 2020.

\bibitem{lenarduzzi2021critical}
V.~Lenarduzzi, S.~Lujan, N.~Saarimaki, and F.~Palomba.
\newblock A critical comparison on six static analysis tools: Detection,
  agreement, and precision, 2021.

\bibitem{lenarduzzi2019does}
V.~Lenarduzzi, V.~Nikkola, N.~Saarimäki, and D.~Taibi.
\newblock Does code quality affect pull request acceptance? an empirical study,
  2019.

\bibitem{10.1007/BFb0026666}
D.~D. Lewis.
\newblock Naive (bayes) at forty: The independence assumption in information
  retrieval.
\newblock In C.~N{\'e}dellec and C.~Rouveirol, editors, {\em Machine Learning:
  ECML-98}, pages 4--15, Berlin, Heidelberg, 1998. Springer Berlin Heidelberg.

\bibitem{lipow1982number}
M.~Lipow.
\newblock Number of faults per line of code.
\newblock {\em IEEE Transactions on software Engineering}, (4):437--439, 1982.

\bibitem{lomio2021fault}
F.~Lomio, S.~Moreschini, and V.~Lenarduzzi.
\newblock Fault prediction based on software metrics and sonarqube rules.
  machine or deep learning?, 2021.

\bibitem{mccabe76}
T.~McCabe.
\newblock A complexity measure.
\newblock {\em IEEE Trans.\ Softw.\ Eng.}, SE-2(4):308--320, Dec 1976.

\bibitem{menzies2010defect}
T.~Menzies, Z.~Milton, B.~Turhan, B.~Cukic, Y.~Jiang, and A.~Bener.
\newblock Defect prediction from static code features: current results,
  limitations, new approaches.
\newblock {\em Automated Software Engineering}, 17(4):375--407, 2010.

\bibitem{7158503}
R.~{Mo}, Y.~{Cai}, R.~{Kazman}, and L.~{Xiao}.
\newblock Hotspot patterns: The formal definition and automatic detection of
  architecture smells.
\newblock In {\em 2015 12th Working IEEE/IFIP Conference on Software
  Architecture}, pages 51--60, May 2015.

\bibitem{10.1109/TSE.2009.50}
N.~Moha, Y.-G. Gueheneuc, L.~Duchien, and A.-F. Le~Meur.
\newblock Decor: A method for the specification and detection of code and
  design smells.
\newblock {\em IEEE Trans. Softw. Eng.}, 36(1):20–36, Jan 2010.

\bibitem{47853}
E.~Murphy-Hill, C.~Jaspan, C.~Sadowski, D.~C. Shepherd, M.~Phillips, C.~Winter,
  A.~K. Dolan, E.~K. Smith, and M.~A. Jorde.
\newblock What predicts software developers’ productivity?
\newblock {\em Transactions on Software Engineering}, 2019.

\bibitem{nagappan05}
N.~{Nagappan} and T.~{Ball}.
\newblock Static analysis tools as early indicators of pre-release defect
  density.
\newblock In {\em Proceedings. 27th International Conference on Software
  Engineering, 2005. ICSE 2005.}, pages 580--586, 2005.

\bibitem{10.1145/42372.42379}
B.~A. Nejmeh.
\newblock Npath: A measure of execution path complexity and its applications.
\newblock {\em Commun. ACM}, 31(2):188–200, Feb 1988.

\bibitem{Niedermayr18Trivial}
R.~Niedermayr, T.~Röhm, and S.~Wagner.
\newblock Too trivial to test? an inverse view on defect prediction to identify
  methods with low fault risk.
\newblock 10 2018.

\bibitem{10.1145/3338501.3357374}
R.~Oak, M.~Du, D.~Yan, H.~Takawale, and I.~Amit.
\newblock Malware detection on highly imbalanced data through sequence
  modeling.
\newblock In {\em Proceedings of the 12th ACM Workshop on Artificial
  Intelligence and Security}, AISec’19, page 37–48, New York, NY, USA,
  2019. Association for Computing Machinery.

\bibitem{Oliveira2020TLProd}
E.~Oliveira, E.~Fernandes, I.~Steinmacher, M.~Cristo, T.~Conte, and A.~Garcia.
\newblock Code and commit metrics of developer productivity: a study on team
  leaders perceptions.
\newblock {\em Empirical Software Engineering}, 04 2020.

\bibitem{pearl2009causality}
J.~Pearl.
\newblock {\em Causality}.
\newblock Cambridge university press, 2009.

\bibitem{potdar2014exploratory}
A.~Potdar and E.~Shihab.
\newblock An exploratory study on self-admitted technical debt.
\newblock In {\em 2014 IEEE International Conference on Software Maintenance
  and Evolution}, pages 91--100. IEEE, 2014.

\bibitem{alex2016data}
A.~Ratner, C.~D. Sa, S.~Wu, D.~Selsam, and C.~Ré.
\newblock Data programming: Creating large training sets, quickly, 2016.

\bibitem{LinusRule}
E.~Raymond.
\newblock The cathedral and the bazaar.
\newblock {\em First Monday}, 3(3), 1998.

\bibitem{real1996probabilistic}
R.~Real and J.~M. Vargas.
\newblock The probabilistic basis of jaccard's index of similarity.
\newblock {\em Systematic biology}, 45(3):380--385, 1996.

\bibitem{replication}
Replication.
\newblock Replication package https://figshare.com/s/f7f0301586cee99d7a9d,
  March 2022.

\bibitem{10.1145/1368088.1368135}
J.~R. Ruthruff, J.~Penix, J.~D. Morgenthaler, S.~Elbaum, and G.~Rothermel.
\newblock Predicting accurate and actionable static analysis warnings: An
  experimental approach.
\newblock In {\em Proceedings of the 30th International Conference on Software
  Engineering}, ICSE ’08, page 341–350, New York, NY, USA, 2008.
  Association for Computing Machinery.

\bibitem{5770619}
H.~{Shen}, J.~{Fang}, and J.~{Zhao}.
\newblock Efindbugs: Effective error ranking for findbugs.
\newblock In {\em 2011 Fourth IEEE International Conference on Software
  Testing, Verification and Validation}, pages 299--308, 2011.

\bibitem{shi2020selective}
X.~Shi, W.~Miao, and E.~T. Tchetgen.
\newblock A selective review of negative control methods in epidemiology.
\newblock {\em Current epidemiology reports}, 7(4):190--202, 2020.

\bibitem{swanson2015definition}
S.~A. Swanson, M.~Miller, J.~M. Robins, and M.~A. Hern{\'a}n.
\newblock Definition and evaluation of the monotonicity condition for
  preference-based instruments.
\newblock {\em Epidemiology (Cambridge, Mass.)}, 26(3):414, 2015.

\bibitem{Tosun11Ai}
A.~Tosun, A.~Bener, and R.~Kale.
\newblock Ai-based software defect predictors: Applications and benefits in a
  case study.
\newblock {\em AI Magazine}, 32:57--68, 06 2011.

\bibitem{trautsch2020longitudinal}
A.~Trautsch, S.~Herbold, and J.~Grabowski.
\newblock A longitudinal study of static analysis warning evolution and the
  effects of pmd on software quality in apache open source projects.
\newblock {\em Empirical Software Engineering}, 25(6):5137--5192, 2020.

\bibitem{1173068}
E.~{van Emden} and L.~{Moonen}.
\newblock Java quality assurance by detecting code smells.
\newblock In {\em Ninth Working Conference on Reverse Engineering, 2002.
  Proceedings.}, pages 97--106, Nov 2002.

\bibitem{van2007experimental}
J.~Van~Hulse, T.~M. Khoshgoftaar, and A.~Napolitano.
\newblock Experimental perspectives on learning from imbalanced data.
\newblock In {\em Proceedings of the 24th international conference on Machine
  learning}, pages 935--942, 2007.

\bibitem{vandenberg1966contributions}
S.~G. Vandenberg.
\newblock Contributions of twin research to psychology.
\newblock {\em Psychological Bulletin}, 66(5):327, 1966.

\bibitem{vapnik2013nature}
V.~Vapnik.
\newblock {\em The nature of statistical learning theory}.
\newblock Springer science \& business media, 2013.

\bibitem{doi:10.1137/1116025}
V.~Vapnik and A.~Chervonenkis.
\newblock On the uniform convergence of relative frequencies of events to their
  probabilities.
\newblock {\em Theory of Probability \& Its Applications}, 16(2):264--280,
  1971.

\bibitem{vassallo2020developers}
C.~Vassallo, S.~Panichella, F.~Palomba, S.~Proksch, H.~C. Gall, and A.~Zaidman.
\newblock How developers engage with static analysis tools in different
  contexts.
\newblock {\em Empirical Software Engineering}, 25(2):1419--1457, 2020.

\bibitem{Walkinshaw:2018:FRD:3239235.3239244}
N.~Walkinshaw and L.~Minku.
\newblock Are 20\% of files responsible for 80\% of defects?
\newblock In {\em Proceedings of the 12th ACM/IEEE International Symposium on
  Empirical Software Engineering and Measurement}, ESEM '18, pages 2:1--2:10,
  New York, NY, USA, 2018. ACM.

\bibitem{10.5555/948835.948838}
M.~Zaidman.
\newblock Teaching defensive programming in java.
\newblock {\em J. Comput. Sci. Coll.}, 19(3):33–43, Jan 2004.

\bibitem{zampetti2022using}
F.~Zampetti, S.~Mudbhari, V.~Arnaoudova, M.~Di~Penta, S.~Panichella, and
  G.~Antoniol.
\newblock Using code reviews to automatically configure static analysis tools.
\newblock {\em Empirical Software Engineering}, 27(1):1--30, 2022.

\bibitem{zampetti2017open}
F.~Zampetti, S.~Scalabrino, R.~Oliveto, G.~Canfora, and M.~Di~Penta.
\newblock How open source projects use static code analysis tools in continuous
  integration pipelines.
\newblock In {\em 2017 IEEE/ACM 14th International Conference on Mining
  Software Repositories (MSR)}, pages 334--344. IEEE, 2017.

\end{thebibliography}

\end{document}